\documentclass[letterpaper,amsmath,
  amssymb,aps,prd,nofootinbib,
singlecolumn,
superscriptaddress,altaffilsymbol,12pt]{revtex4-1}

\usepackage{amsmath}
\usepackage{amsfonts}
\usepackage{amssymb}
\usepackage{graphicx}
\usepackage{psfrag}
\usepackage[hang]{footmisc}
\usepackage{setspace}
\usepackage{caption}
\usepackage[dvipsnames]{xcolor}
\newcommand{\beq}{\begin{equation}}
\newcommand{\eeq}{\end{equation}}
\newcommand{\bea}{\begin{eqnarray}}
\newcommand{\eea}{\end{eqnarray}}
\newcommand{\beqn}{\begin{equation*}}
\newcommand{\eeqn}{\end{equation*}}
\newcommand{\bean}{\begin{eqnarray*}}
\newcommand{\eean}{\end{eqnarray*}}

\newcommand*{\cref}[1]{Chapter~\ref{#1}}

\usepackage{graphicx}
\usepackage{dcolumn}
\usepackage{psfrag} 
\usepackage{booktabs}

\begin{document}
%\begin{flushright} {\footnotesize IC/2007/001
%\\ HUTP-07/A0002}  \end{flushright} 
\title{Precise determination of the inflationary epoch and constraints for reheating}% Force line breaks with \\

 \author{Gabriel Germ\'an}\email{e-mail: gabriel@icf.unam.mx}
 \affiliation{Instituto de Ciencias F\'{\i}sicas, Universidad Nacional Aut\'onoma de M\'exico, Cuernavaca, Morelos, 62210, Mexico}
%\date{\today}
\begin{abstract}
We present a simple formula that allows to calculate the value of the inflaton field, denoted by $\phi$, at the scale with wavenumber mode $k$. In the extreme case of instantaneous reheating $\phi_k$ is calculated exactly and all inflationary observables and quantities of interest follow. This formula, together with the fact that the scale factor $a_p$ at the pivot scale wavenumber $k_p=0.05/Mpc$ lies in the radiation era, allows the development of a diagrammatic approach to study the evolution of the universe. This scheme is complementary to the usual analytical method and  some interesting results, independent of the model of inflation, can be obtained. As a concrete application of the ideas developed here we discuss them with some detail using the Starobinsky model of inflation.

\end{abstract}

\maketitle
%%%%%%%%%%%%%%%%%%%%%%%%%%%%%%%%
\section {\bf Introduction}\label{Intro}

There have been several recent attempts to establish constraints on inflationary models  \cite{Guth:1980zm},  \cite{Linde:1984ir},  \cite{Lyth:1998xn},  \cite{Martin:2018ycu},  particularly during the epochs of inflation and reheating, with varying degrees of success \cite{Liddle:2003as}  -  \cite{Ji:2019gfy}. For reviews on reheating see e.g.,  \cite{Bassett:2005xm}, \cite{Allahverdi:2010xz},  \cite{Amin:2014eta}. 
Here, we present a simple approach where the inflationary epoch is precisely determined by constructing an equation for $\phi_k$, the value of the inflation field at the comoving Hubble scale with wavenumber mode $k\equiv a_kH_k$ which we set equal to the pivot scale $k_p$. We use $k_p=\frac{0.05}{Mpc}$ where most parameter values are reported, in particular by the Planck collaboration \cite{Aghanim:2018eyx}, \cite{Akrami:2018odb}. Once we have determined  $\phi_k$ all inflationary quantities of interest follow. The outline of the paper is as follows: in sections \ref{INF} and \ref{REH} we establish general results which then apply to a particular model in Section~\ref{STA}.  Section~\ref{INF} provides a brief discussion on the obtention of the $\phi_k$-equation an establish it in Eq.~\eqref{EQ}. Section~\ref{REH} studies the reheating epoch where formulas for the number of e-folds during reheating and during radiation domination are given as functions of the reheating temperature. Also bounds for these quantities are found for the minimal reheating temperature $T_{re}\approx 10\,MeV$ as required by nucleosynthesis.
Section~\ref{STA} contains a study of the Starobinsky model along the lines described above. This is done in the $\omega_{re}=0$ case, where $\omega_{re}$ is the equation of state parameter (EoS) during reheating. We also compare the diagrammatic approach with the usual study of analytical expressions and show how both procedures give essentially the same results and complement each other. In Section~\ref{INS} we solve exactly the case of instantaneous reheating for the Starobinsky model. Finally, Section~\ref{CON} contains the main conclusions of the paper.
%%%%%%%%%%%%%%%%%%%%%%%%%%%%%%%%
%%%%%%%%%%%%%%%%%%%%%%%%%%%%%%%%
\section {\bf The inflationary epoch}\label{INF}
\begin{figure}[tb]
\captionsetup{justification=raggedright,singlelinecheck=false}
\par
\includegraphics[trim = 0mm  0mm 1mm 1mm, clip, width=13.6cm, height=8cm]{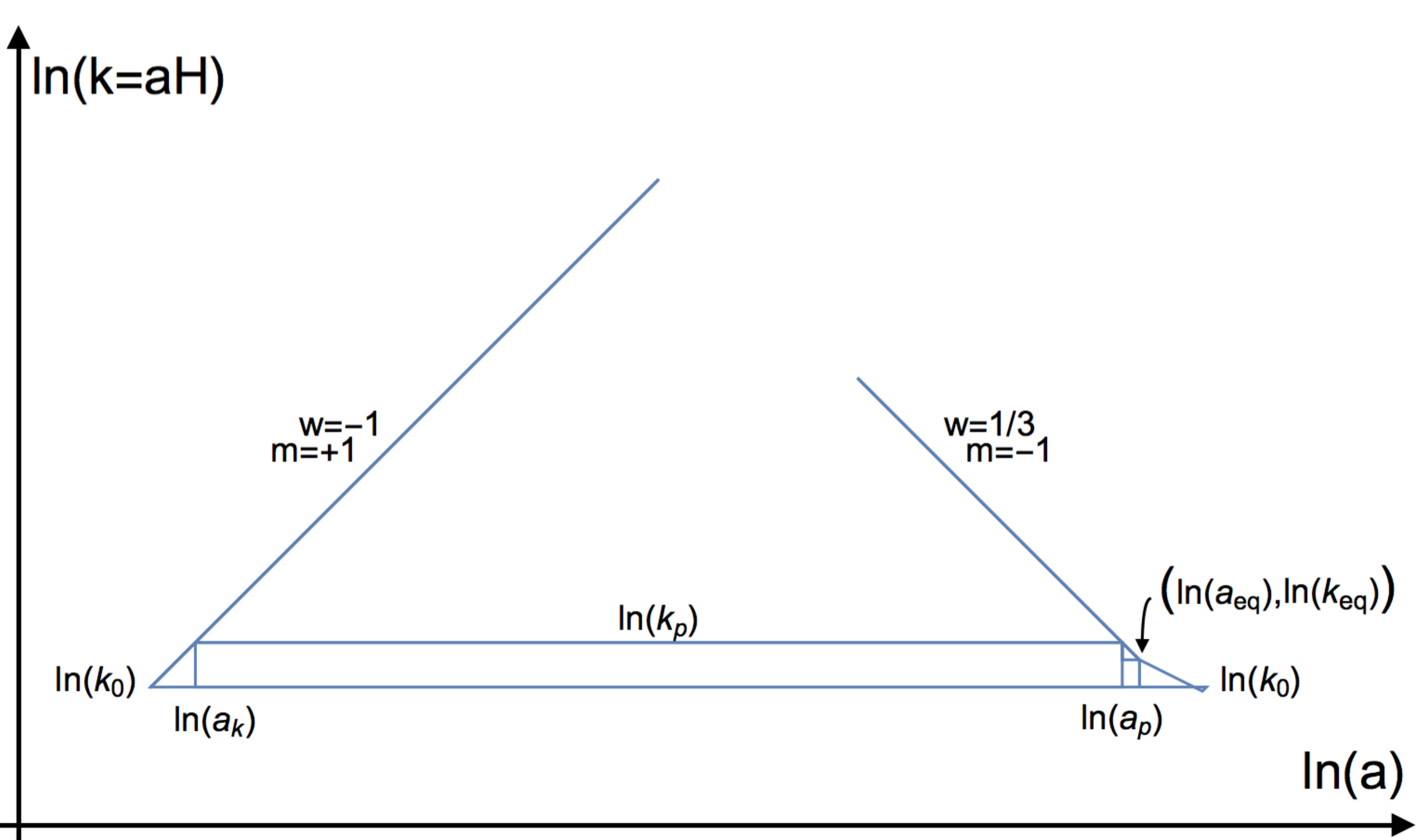}
\caption{\small  The radiation line (EoS $\omega_{re} = 1/3$ and slope $m = -1$) is fixed on the 
pivot point with coordinates $\left(\ln(a_p), \ln(k_p)\right)$, where $a_p$ is the scale factor determined on the pivot scale. The inflation line (EoS $\omega_{re} = -1$ and slope $m = +1$) is fixed by determining the total number of e-folds $N_{ke} + N_{ep}=\ln(a_p)-\ln(a_k)$ along the line $\ln(k_p)$
(see discussion in Section~\ref{STA}). These fixed lines constitute a framework on which measurements can be carried out and the number of e-folds determined between any pair of points on the diagram. Once the diagram is connected by a reheating line joining the inflation with the radiation line (see Fig.~\ref{DSconectado}) a diagram of the evolution of the universe has been constructed where measurements of intervals and location of points can be carried out with accuracy.}
\label{DSincompleto}
\end{figure}

We work in Planck mass units, where $M_{pl}=2.4357\times 10^{18} GeV$ and set $M_{pl}=1,$ the pivot scale wavenumber $k_p\equiv a_pH_p=0.05\frac{1}{Mpc}$, used in particular by the Planck collaboration, becomes a dimensionless number given by $k_p= 1.3105\times 10^{-58}$. This can be compared with $k_0\equiv a_0H_0 = 8.7426\times 10^{-61}h$.
To find the value of $a_p$ and from there the number of e-folds $N_{p0}\equiv\ln\frac{a_{0}}{a_{p}}$ from $a_p$ up to the present at $a_0$ we solve the Friedmann equation for $a_p$
\begin{equation}
k_p=H_0\sqrt{\frac{\Omega_{md,0}}{a_p}+\frac{\Omega_{rd,0}}{a_p^2}+\Omega_{de}a_p^2}\;, 
 \label{Frid}
\end{equation}
where $\Omega_{md,0}=0.315,$ $\Omega_{rd,0}=5.443\times 10^{-5},$ and $\Omega_{de}=0.685.$ 
To calculate $a_p$ we have to specify $h$ for the Hubble parameter $H_0$ at the present time, we take the value $h=0.674$ given  by Planck. The solution of Eq.~\eqref{Frid} is $a_p=3.6512\times 10^{-5}$ from where we get $N_{p0}=10.22$ for the number of e-folds from $a_p$ to $a_0$. Thus, $a_p < a_{eq}\approx 2.97\times 10^{-4}$ where $a_{eq}$ is the scale factor at matter-radiation equality and  $k_p$ is inside the radiation dominated era. This allows to fix the radiation line of EoS $\omega=1/3$ and slope $m=-1$ passing through the point $\left(\ln(a_p),\ln(k_p)\right)$
(see Fig.~\ref{DSincompleto} and reference \cite{German:2020kdp} for a thorough discussion of the diagrammatic approach).

Defining $N_{ke}=\ln\frac{a_e}{a_k}$ as the number of e-folds from $a_k$ to the end of inflation at $a_e$, $N_{re}=\ln\frac{a_{re}}{a_{e}}$ the number of e-folds during reheating , $N_{rd}=\ln\frac{a_{eq}}{a_{r}}$ during radiation and $N_{eq0}=\ln\frac{a_{0}}{a_{eq}}$ from matter-radiation equality to the present, it is easy to show that the number of e-folds from reheating {\it plus} the radiation dominated epochs can be written as 
\begin{equation}
N_{re}+N_{rd}=N_{ke}+N_{re}+N_{rd}-N_{ke}=\ln[\frac{a_{eq}H_k}{k}]-N_{ke}\;, 
 \label{re+rd}
\end{equation}
\\
we see that the r.h.s only depends on $\phi_k$, the value of the inflaton at $k$ (and also of parameters of the model, if any). The equation which determines $\phi_k$  is
\begin{equation}
N_{re} + N_{rd} -N_{peq} = N_{ep}\;, 
\label{enes}
\end{equation}
where $N_{peq}\equiv \ln(\frac{a_{eq}}{a_p})\approx 2.1$ is the number of e-folds from the pivot scale factor $a_p$ to mater-radiation equality at $a_{eq}$ and $N_{ep}\equiv \ln(\frac{a_p}{a_e})$ is the number of e-folds from the end of inflation to $a_p$ at the pivot scale. This equation can also be written as $k=k_p$ or
\begin{equation}
\ln[\frac{a_{p}H_k}{k_p}]=N_{ke}+N_{ep}\;.
\label{EQ}
\end{equation}
Solving Eq.~\eqref{EQ} for a given $N_{ep}$ requires specifying a model of inflation; $H_k$ and $N_{ke}$ are model dependent quantities.
Thus, after finding $\phi_k$,  we can proceed to determine all inflationary quantities like the scale of inflation, Hubble parameter $H_k$, tensor-to-scalar ratio $r$, spectral index $n_s$, running $\alpha$, etc.  Notice how the determination of $\phi_k$ requires not only the knowledge of the present universe through quantities like $\Omega_{i,0}$ and $H_0$ but also of the early universe through the scalar power spectrum amplitude given here by $A_s$ and contained in $H_k=\sqrt{8\pi^2 \epsilon_k A_s}$ where $\epsilon_k$ is the slow-roll parameter $\epsilon\equiv\frac{1}{2}\left(\frac{V_{\phi}}{V}\right)^2$ at $\phi_k$.

An equivalent way of obtaining Eq.~\eqref{EQ} is by connecting the epoch when the scale of wavenumber $k$ left the horizon during inflation to the pivot scale at $k_p$ where we measure the horizon reentry of precisely the same scale. This can be expressed by 
\begin{equation}
\ln\left(\frac{a_{p}}{a_k}\right)=N_{ke}+N_{ep}\;.
\label{eq1}
\end{equation}
Multiplying the term inside the parenthesis above and below by $H_k$ and setting $a_k H_k \equiv k = k_p\equiv a_p H_{p}$ we get again Eq.~\eqref{EQ}.

In general we can numerically solve Eq.~\eqref{EQ} by requiring agreement with the data e.g., for an spectral index $n_s=0.9649\pm 0.0042$ and tensor-to-scalar index $r < 0.063$; this fixes the range of values $N_{ep}$ can take. We can parameterize $N_{ep}$ by $N_{ep}\equiv\alpha N_{ke}$  where $\alpha$ is a positive parameter around $1$ such that when $\alpha = 1$ $N_{ep} = N_{ke}$ and the diagram is perfectly symmetric, from $\ln (k_p)$ upwards, around an axis passing through the vertex $V$ (see e.g., Fig.~\ref{DSparalelas}). This case corresponds to instant reheating for any $\omega_{re}$ (see discussion in Section~\ref{INS}). For $\alpha> 1$ then $\omega_{re} < 1/3$ and $N_{ep} > N_{ke}$ and when $\alpha < 1$ then $\omega_{re} > 1/3$ with $N_{ep} < N_{ke}$. In Starobinsky model the range $0.9607 < n_s < 0.9691$ implies $1.275 > \alpha > 0.774$.
%%%%%%%%%%%%%%%%%%%%%%%%%%%%%%%%%%%%%%%%%
%%%%%%%%%%%%%%%%%%%%%%%%%%%%%%%%%%%%%%%%%
%%%%%%%%%%%%%%%%%%%%%%%%%%%%%%%%%%%%%%%%%
\section {\bf The reheating epoch}\label{REH}

Having specified a formula from where $\phi_k$ can, in principle, be obtained and from there all relevant quantities characteristic of the inflationary epoch we now turn to the reheating era. Assuming a constant equation of state parameter for any $\omega_{re}$ during reheating the fluid equation gives $\rho \varpropto a^{-3(1+\omega)}$ from where the number of e-folds during reheating follows
\begin{equation}
N_{re}\equiv\ln\left(\frac{a_{r}}{a_{e}}\right) = [3(1+\omega_{re})]^{-1}\ln[\frac{\rho_e}{\rho_{re}}]\;,
\label{NRE1}
\end{equation}
where $\rho_e$ denotes the energy density at the end of inflation and $\rho_{re}$ the energy density at the end of reheating. This quantity is given by $\rho_{re} = \frac{\pi^2 g_{re}}{30} T_{re}^4$ and $g_{re}$ is the number of degrees of freedom of relativistic species at the end of reheating. Assuming entropy conservation after reheating 
\begin{equation}
g_{s,re}T_{re}^3=\left(\frac{a_0}{a_{eq}}\right)^3\left(\frac{a_{eq}}{a_{r}}\right)^3\left(2 T_0^3+6\times \frac{7}{8}T_{\nu,0}^3\right)\;,
\label{entropy}
\end{equation}
where $T_0=2.725 K$ and the neutrino temperature is $T_{\nu,0}=(4/11)^{1/3} T_0$. The number of e-folds during radiation domination $N_{rd}\equiv\ln\frac{a_{eq}}{a_{r}}$ follows from Eqs.~\eqref{NRE1} and \eqref{entropy}
\begin{equation}
N_{rd}= -\frac{3(1+\omega_{re})}{4}N_{re}+\frac{1}{4} \ln[\frac{30}{g_{re} \pi^2}] +\frac{1}{4} \ln[\frac{\rho_e}{T_0^4}] +\frac{1}{3} \ln[\frac{11 g_{s,re}}{43}]+\ln[\frac{a_{eq}}{a_0}]\;.
 \label{NRD}
\end{equation}
Combining Eqs.~\eqref{re+rd} and \eqref{NRD} we get an expression for the number of e-folds during reheating 
\begin{equation}
N_{re}= \frac{4}{1-3\, \omega_{re}}\left(-N_{ke}-\frac{1}{3} \ln[\frac{11 g_{s,re}}{43}]-\frac{1}{4} \ln[\frac{30}{\pi^2 g_{re} } ] -\ln[\frac{ \rho^{1/4}_e k}{H_k\, a_0 T_0} ]\right)\;. 
 \label{NRE}
\end{equation}
It is convenient to rewrite this equation in the form
\begin{equation}
N_{re}=\frac{4}{1-3\, \omega_{re}}\bar{N}_{re}\;, 
 \label{NRE2}
\end{equation}
where $\bar{N}_{re}$ is the term in the brackets of Eq.~\eqref{NRE} and is independent of $\omega_{re}$.
A final quantity of physical relevance is the thermalization temperature at the end of the reheating phase
\beq
\label{TRE}
T_{re}=\left( \frac{30\, \rho_e}{\pi^2 g_{re}} \right)^{1/4}\, e^{-\frac{3}{4}(1+\omega_{re})N_{re}}\,.
\eeq
\noindent 
This is a function of the number of $e$-folds during reheating. It can also be written as an equation for the parameter $\omega_{re}$, using  Eq.~\eqref{NRE2}
\begin{equation}
\omega_{re}=\frac{1}{3}+\frac{4 \bar{N}_{re}}{3\left(-\bar{N}_{re}+\frac{1}{4}\ln[\frac{\pi^2\,g_{re}} {30\rho_e}\, {T_{re}}^4]\right)}\;, 
 \label{w}
\end{equation}
from here we can rewrite the equations for $N_{re}$ and $N_{rd}$ as functions of $T_{re}$ and $n_s$ and of $T_{re}$, respectively
\begin{equation}
N_{re}= \bar{N}_{re}-\frac{1}{4}\ln[\frac{\pi^2\,g_{re}} {30\rho_e}] - \ln[T_{re}]\;, 
 \label{NRET}
\end{equation}
\begin{equation}
N_{rd}= \ln[\frac{a_{eq}}{a_0\,T_0}]+\frac{1}{3} \ln[\frac{11 g_{s,re}}{43}]+\ln[T_{re}]\;, 
 \label{NRDT}
\end{equation}
from where we see that $N_{re}+N_{rd}$ is $T_{re}$ independent, equivalently $\omega_{re}$ independent. As shown in Eq.~\eqref{re+rd} the sum $N_{re}+N_{rd}$ only depends on $\phi_k$, the value of the inflaton at $k$ (and also of parameters of the model, if any).
%%%%%%%%%%%%%%%%%%%%%%%%%%%%%%%%%%%%%%%%%
%%%%%%%%%%%%%%%%%%%%%%%%%%%%%%%%%%%%%%%%%
\section {\bf The Starobinsky model: the diagrammatic and the analytical views }\label{STA}
The potential of the Starobinsky model \cite{Starobinsky:1980te,Mukhanov:1981xt,Starobinsky:1983zz} is given by \cite{Whitt:1984pd}:
\beq
\label{staropot}
V= V_0 \left(1- e^{-\sqrt{\frac{2}{3}}\phi} \right)^2.
\eeq
From here we calculate the number of e-foldings from  $\phi_k$ up to the end of inflation
\beq
\label{Nke}
N_{ke} = -\int_{\phi_k}^{\phi_e}\frac{V}{V'}d\phi = \frac {1}{4} \left( 3e^{\sqrt{\frac{2}{3}}\phi_k} -\sqrt{6} \,\phi_k \right)-\frac {1}{4} \left( 3e^{\sqrt{\frac{2}{3}}\phi_e} -\sqrt{6} \,\phi_e \right),
\eeq
where the end of inflation is given by the solution to the equation $\epsilon\equiv \frac{1}{2}\left(\frac{V_{\phi}}{V}\right)^2 =1$ at $\phi_e$: $\phi_e= \sqrt{\frac {3}{2}}\ln\left(1+\frac{2}{\sqrt{3}}\right)$. The Hubble function is 
\beq
\label{Hk}
H_k=\sqrt{8\pi^2 \epsilon_k A_s}=\sqrt{\frac{32 A_s}{3}}\frac{\pi}{e^{\sqrt{\frac{2}{3}}\phi_k}-1}\,,
\eeq
where $\epsilon_k$ is the slow-roll parameter $\epsilon$ at $\phi_k$ and the scalar power spectrum amplitude is $A_s=2.0991 \times 10^{-9}$. 
From the equation for the spectral index $n_s=1+2\eta-6\epsilon$ the solution for $\phi_k$  in terms of $n_s$ is
\beq
\label{fikf}
\phi_k= \sqrt{\frac {3}{2}}\ln      \left(\frac{7-3n_s+4\sqrt{4-3n_s}}{3(1-n_s)}  \right).
\eeq
\noindent
In this section we also study how to fix a diagrammatic framework using Starobinsky model of inflation as a working example. First we notice that  in a diagram lines are associated with an EoS $\omega$ as well as with a slope $m$. The relation between them is given by
\begin{equation}
m=-\frac{1+3\omega}{2}\;,
\label{m}
\end{equation}
Thus, the line representing radiation is given by an with EoS $\omega=1/3$ and slope $m=-1$ and fixed by the scale factor $a_p$ lying in the radiation era.
The inflation line (EoS $\omega=-1$ and slope $m=+1$) is fixed by finding the total number of e-folds $N_{ke}+N_{ep}$ from $a_k$ to $a_p$ represented by the horizontal $\ln(k_p)$ line from $\ln a_k$ to $\ln a_p$ in Fig.~\ref{DSincompleto}. 
In the diagrammatic approach the Hubble function $H$ is assumed  constant during inflation (de Sitter universe)
but for specific models $H_k$ is not a constant during inflation. 
To assign a reasonable value to $H$ we can solve the l.h.s of  Eqs.~\eqref{EQ} for the range given by Planck to the spectral index $0.9607 < n_s < 0.9691$ obtaining
\begin{equation}
111.15 < \ln[\frac{a_{p}H_k}{k_p}] < 110.92\;.
\label{EQparticular}
\end{equation}
Thus, a reasonable value to fix the ``distance" from $\ln a_k$ to $\ln a_p$ would be $\ln[\frac{a_{p}H_k}{k_p}]_{mean}=111.04$. As we see in the Starobinsky model, with the assigned $\ln[\frac{a_{p}H_k}{k_p}]_{mean}=111.04$ maximum departures from constant $H$ amount to a $0.1\%$ of the total number of e-folds from $a_k$ to $a_p$; a mere 0.1 e-fold in this case, this particular example shows the good nature of the approximation.
Thus, the construction described above fixes completely the frame where the diagrammatic approach is based.
Dividing this number (111.04) by the length of the $\ln(a_p)-\ln(a_k)$ segment we find the total number of e-folds per unit length for the Starobinsky model at $k=k_p$. 
In this article we do not make bound estimates with the diagrams but only use them for illustrative purposes (see   \cite{German:2020kdp} for a quantitative, model independent use of the diagrammatic approach).
From Fig.~\ref{DSincompleto} we see that to determine the history of the evolution of the universe according to the Starobinsky model we have to connect the diagram i.e., to join the inflation line to the radiation line, that (connecting) line will be called the reheating line. Once we get a connected diagram we can proceed to measure the ``distances" between any two points in the diagram \cite{German:2020kdp}. Thus, to connect the diagram we have to specify one point in the inflation line and one point in the radiation line or one point in any line and the slope of the reheating line. 
%%%%%%%%%%%%%%%%%%%%%%%%%%%%%%%%%%%%%%%%%
\begin{figure}[t!]
\captionsetup{justification=raggedright,singlelinecheck=false}
\par
\begin{center}
\includegraphics[trim = 0mm  0mm 1mm 1mm, clip, width=13.6cm, height=8cm]{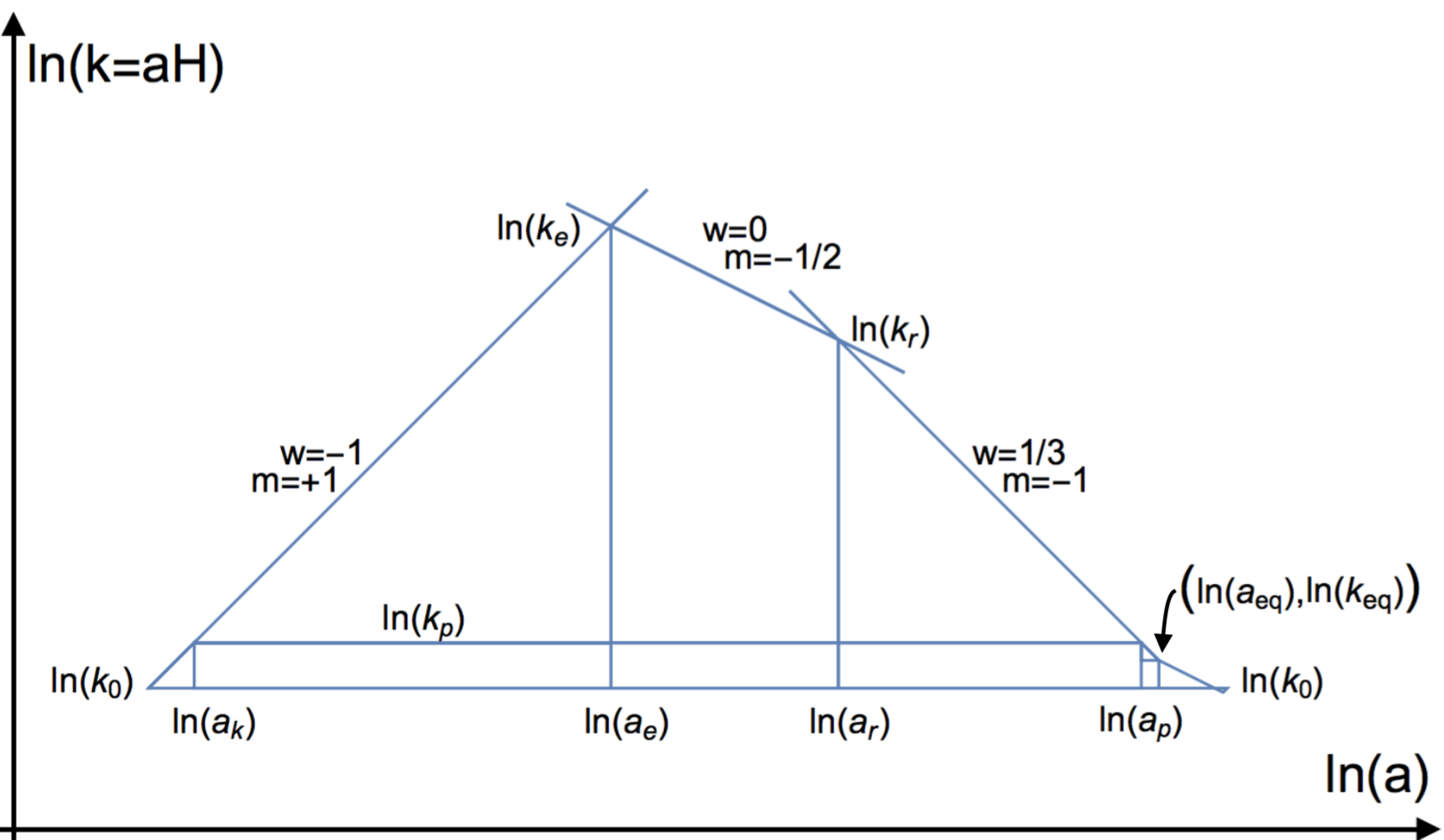}
\end{center}
\caption{The diagram of Fig.~\ref{DSincompleto} is now connected with a reheating line joining the inflation with the radiation line, for the Starobinsky model of inflation. This has been done by using the lower bound of Planck's reported range for the spectral index $n_s=0.9649\pm 0.0042$ and by assuming an EoS $\omega_{re} = 0$ of slope $m = -1/2$. This is a reasonable assumption because the Starobinsky model is well approximated in the vicinity of the origin by a quadratic potential. In any case it is a working example where we can compare the diagrammatic and analytical methods of study. For an spectral index larger than the lower bound $n_s=09607$, the reheating line joining the points $\left(\ln(a_e), \ln(k_e)\right)$ with $\left(\ln(a_r), \ln(k_r)\right)$ will go upwards parallel to itself with the same EoS (see Fig.~\ref{DSparalelas}).}
\label{DSconectado}
\end{figure}
%%%%%%%%%%%%%%%%%%%%%%%%%%%%%%%%%%%%%%%%%
%%%%%%%%%%%%%%%%%%%%%%%%%%%%%%%%%%%%%%%%%
\subsection {\bf The Starobinsky model for the ${\bf \omega=0}$ case }\label{STA1}
As an example we specify a point in the inflation line by calculating the number of e-folds during inflation $N_{ke}$ for Planck's lower limiting value for the spectral index $n_s=0.9607$. From Eqs.~\eqref{Nke} and  \eqref{fikf} we get $N_{ke}=48.86$ at wavenumber $k=k_p$. Because Starobinsky model is well approximated by a quadratic potential near the origin we choose a line of slope $m=-1/2$ corresponding to an EoS $\omega_{re}=0$. Thus, the diagram is now connected as shown in Fig.~\ref{DSconectado} with the following results for the reheating and the radiation periods
obtained by solving Eqs~(\ref{NRE}) and (\ref{NRD}) for $n_s=0.9607$ and  EoS $\omega_{re}=0$
\beq
\label{boundsA}
N_{re}=27.04, \quad N_{rd}=37.35.
\eeq
We solve Eq~(\ref{TRE}) to find the reheat temperature, $T_{re}=4.4\times10^{6}\,GeV$.
If we add together all these results for the number of e-folds we get 48.86+27.04+37.35-2.1=111.15 (we substract 2.1 e-folds because the number of e-folds from $a_p$ to the end of radiation at $a_{eq}$ is $N_{peq}=2.1$). 

\begin{figure}[tb]
\captionsetup{justification=raggedright,singlelinecheck=false}
\par
\includegraphics[trim = 0mm  0mm 1mm 1mm, clip, width=13.6cm, height=8cm]{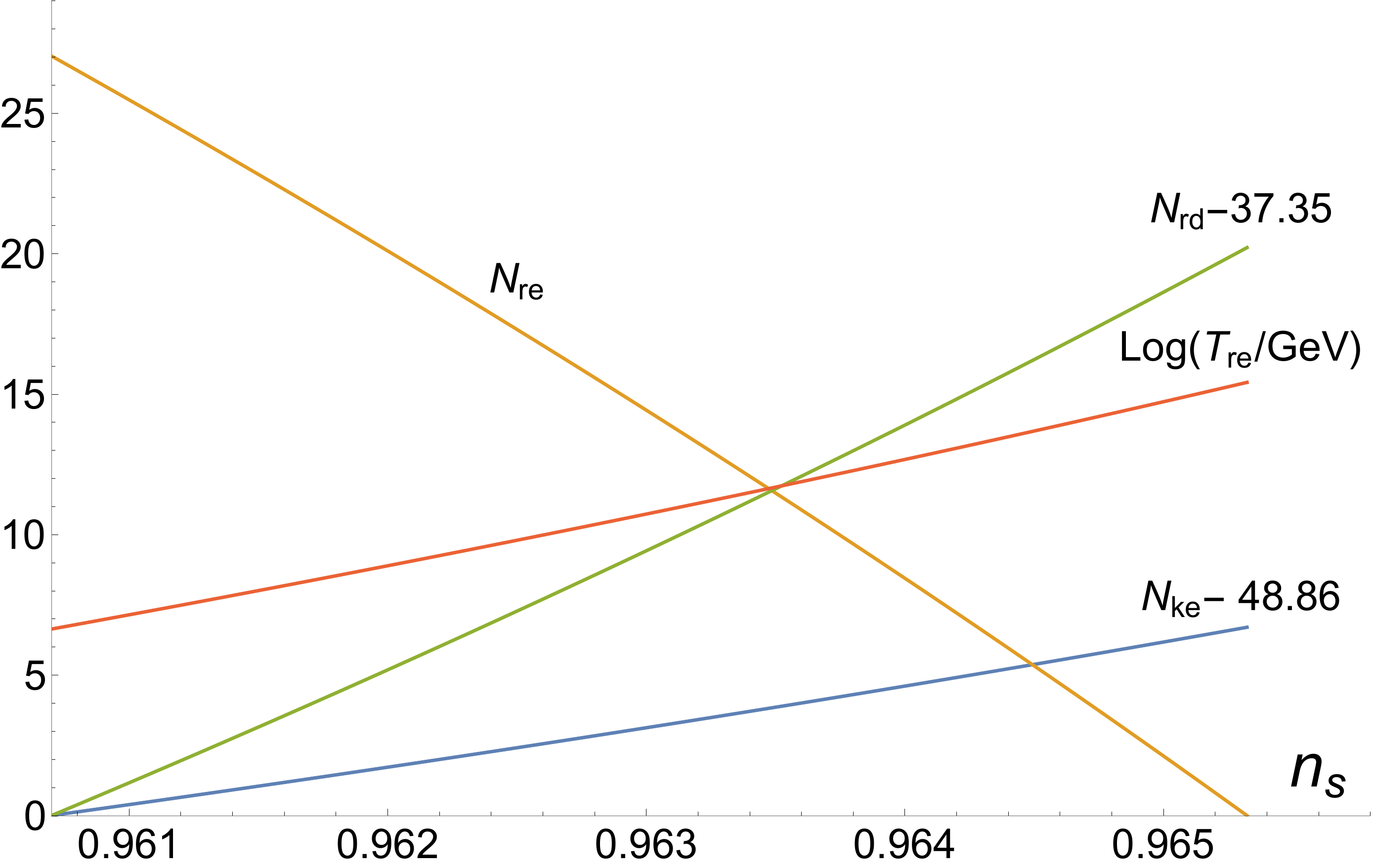}
\caption{\small We plot the number of e-folds during inflation $N_{ke}$ as given by Eq.~\eqref{Nke}, reheating $N_{re}$ as given by Eq.~\eqref{NRE} radiation $N_{rd}$ as given by Eq.~\eqref{NRD} and the $log$ of the reheating temperature $T_{re}$ as given by Eq.~\eqref{TRE}  as functions of the spectral index. The figure shows the precise evolution of each quantity as we move from the lower value of the Planck bound $n_s=0.9607$ to the value of $n_s\approx 0.9653$ where instant reheating occurs ($N_{re}=0$ and $T_{re}=2.6\times 10^{15}\,GeV$ is the maximum temperature), all of this for an EoS $\omega_{re}=0$. Note that each plotted quantity has its own EoS ($\omega=-1$ for the inflation line, $\omega=0$ for reheating, $\omega=1/3$ for radiation, etc.,) but we cannot $see$ this in the plot nor we can $see$ the full plot i.e., how is that all these epochs are connected to each other.  To advance in this direction we can turn to Fig.~\ref{DSparalelas}.}
\label{DSdai}
\end{figure}

\begin{figure}[t!]
\captionsetup{justification=raggedright,singlelinecheck=false}
\par
\begin{center}
\includegraphics[trim = 0mm  0mm 1mm 1mm, clip, width=13.6cm, height=8cm]{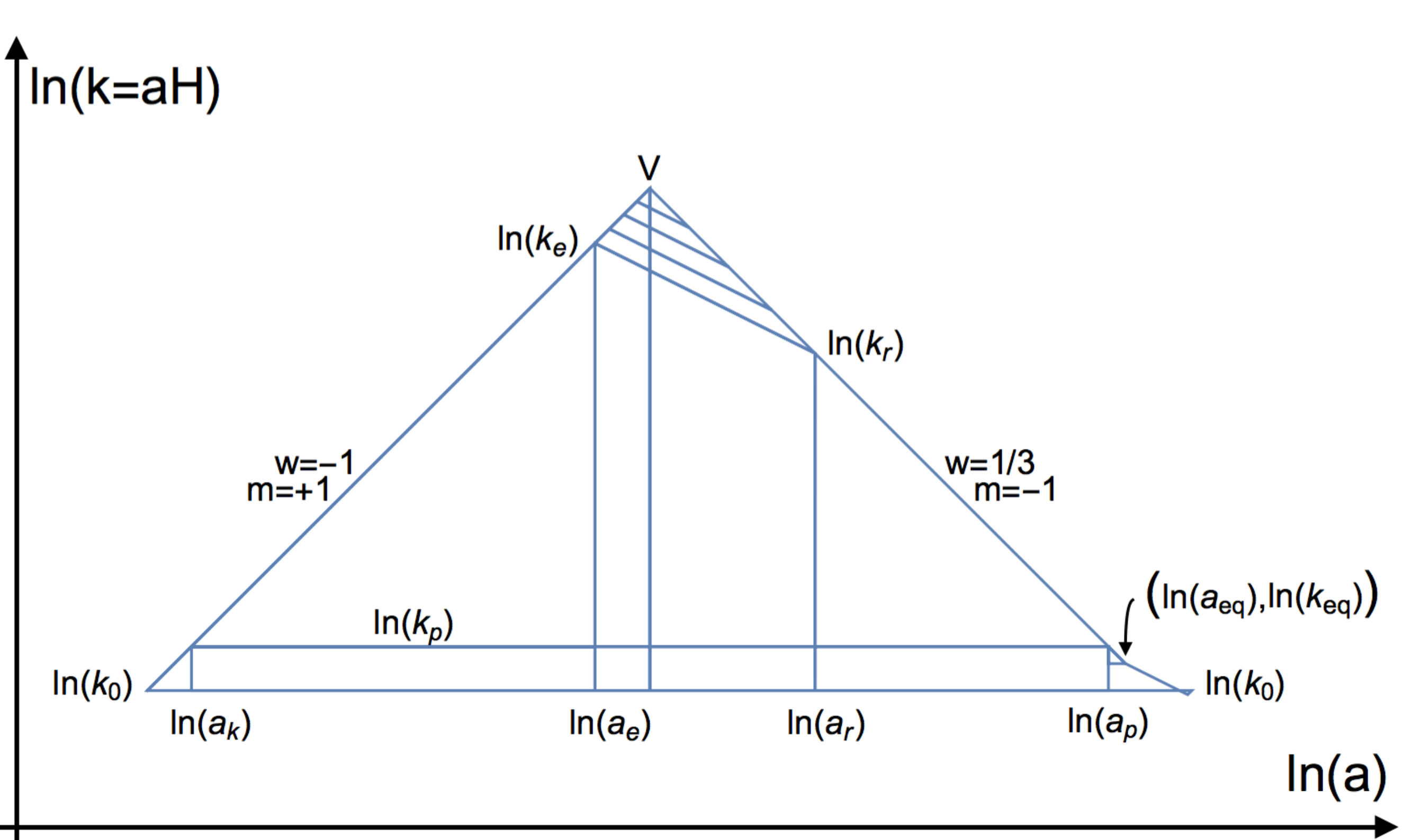}
\end{center}
\caption{The line connecting the points $\left(\ln(a_e),\ln(k_e)\right)$ and $\left(\ln(a_r),\ln(k_r)\right)$ represents a vertical cut through the reheating and radiation lines in Fig.~\ref{DSdai} for the lower bound $n_s=0.9607$ (the ``y" axis). The following line upwards parallel to the previous one and with the same EoS also represents a vertical cut in Fig.~\ref{DSdai} for a sligthly higher value of the spectral index, and so on. We can think of the figure above as a succession of vertical cuts in Fig.~\ref{DSdai} each with a higher value of $n_s$, $N_{ke}$, $N_{rd}$ and $T_{re}$ but a lower value for $N_{re}$. As the parallel lines reach the vertex at $V$ the number of e-folds during reheating goes to zero and the temperature reaches a maximum value (see Eq.~\eqref{TRE}). This is what we usually call instantaneous reheating. Thus, in the diagram above we can see at once a succession of an infinite number of universes (infinite number of parallel lines) with the diverse substances which drive the expansion of the universe represented by connected lines of varios slopes (various EoS). 
This behaviour is the same for any other $\omega_{re}$ and not specific to the $\omega_{re}=0$ case (not specific for a line of slope $m=-1/2$). 
When $\omega_{re} < 1/3$ the parallel lines (with their corresponding slope) converge at $V$ from the left as shown above and for $\omega_{re} > 1/3$ converge at $V$ from the right, with a completely similar construction. In any case the framework bounded by the inflation line of slope $m=+1$ the radiation line of slope $m=-1$ and the horizontal $\ln(k_p)$ line remain fixed and this is what makes possible the diagrammatic description in a precise and quantifiable way.}
\label{DSparalelas}
\end{figure}

To better understand the usefulness of our diagrams and their relation with the more standard approach we have plotted in Fig.~\ref{DSdai} all the  quantities involved as functions of the spectral index ($N_{ke}$ as given by Eq.~\eqref{Nke}, $N_{re}$ as given by Eq.~\eqref{NRE}, $N_{rd}$ as given by Eq.~\eqref{NRD} and $T_{re}$ as given by Eq.~\eqref{TRE}). The figure shows the precise evolution of each quantity as we move from the lower value of the Planck bound $n_s=0.9607$ to the value of $n_s= 0.9653...$ where instant reheating occurs ($N_{re}=0$ and $T_{re}=2.6\times 10^{15}\,GeV$, the maximum possible temperature), all of this for an EoS $\omega_{re}=0$. Note that each plotted quantity in Fig.~\ref{DSdai} has its own EoS but we cannot $see$ this in the plot and we cannot $see$ the full plot i.e., how is that all epochs are connected?
This we can see with the diagrams, let us consider the plot in Fig.~\ref{DSparalelas}. The line connecting the points $\left(\ln(a_e),\ln(k_e)\right)$ and $\left(\ln(a_r),\ln(k_r)\right)$ represents a vertical cut through the reheating and radiation lines in Fig.~\ref{DSdai} i.e., just one set of points  for $n_s=0.9607$ (one possible universe).

Thus, in Fig.~\ref{DSdai} we see the evolution of {\it vertical cuts} (possible universes) as functions of $n_s$. In the diagram of Fig.~\ref{DSparalelas} we also see this by the evolution of parallel lines (all with EoS  $\omega_{re}=0$ and slope $m=-1/2$) converging into the vertex $V$ at $n_s\approx 0.9653$. These parallel lines go upwards as $n_s$ increases from $n_s=0.9607$, $N_{ke}$ increases from 48.86, $N_{rd}$ increases from 37.59, the temperature increases from $T_{re}=4.4\times10^{6}\,GeV$ but $N_{re}$ $decreases$ from 26.65. At the tip of the figure $N_{ke}=55.6$, $N_{rd}=57.6$, $T_{re}=2.6\times10^{15}\,GeV$ and $N_{re}=0$ (instantaneous reheating), essentially the same values obtained from Fig.~\ref{DSdai} at $n_s= 0.9653...$. Each diagram for each one of the infinite number of parallel lines would represent a possible universe. 
This behaviour is the same for any other $\omega_{re}$ and is not specific to the $\omega_{re}=0$ case. 
When $\omega_{re} < 1/3$ the parallel lines (of a given slope) converge at $V$ from the left as shown in Fig.~\ref{DSparalelas} for the $\omega_{re} =0$ case, and for $\omega_{re} > 1/3$ converge at $V$ from the right, with a completely similar construction. 
In any case we see that the framework bounded by the inflation line of slope $m=+1$ the radiation line of slope $m=-1$ and the horizontal $\ln(k_p)$ line remain fixed and this is what makes possible the diagrammatic description in a precise and quantifiable way. With the analytical description we cannot study the $\omega_{re}=1/3$ case because for $\omega_{re}=1/3$ there is no equation for  $N_{re}$  \cite{Cook:2015vqa}. However the diagrammatic approach can be applied without difficulty. The diagrammatic approach can also describe model independent situations where the analytical equations presented before cannot \cite{German:2020kdp}. We clearly see how the diagrammatic and analytical approaches complement each other very nicely and should be studied together (with the help of Eq.~\eqref{EQ})  for better results.
%%%%%%%%%%%%%%%%%%%%%%%%%%%%%%%%%%%%%%%%%
%%%%%%%%%%%%%%%%%%%%%%%%%%%%%%%%%%%%%%%%%
%%%%%%%%%%%%%%%%%%%%%%%%%%%%%%%%%%%%%%%%%
%%%%%%%%%%%%%%%%%%%%%%%%%%%%%%%%%%%%%%%%%
\subsection {\bf The Starobinsky model for the $T_{re}=10\,MeV$ case }\label{STA2}
\begin{figure}[tb]
\captionsetup{justification=raggedright,singlelinecheck=false}
\par
\includegraphics[trim = 0mm  0mm 1mm 1mm, clip, width=13.6cm, height=8cm]{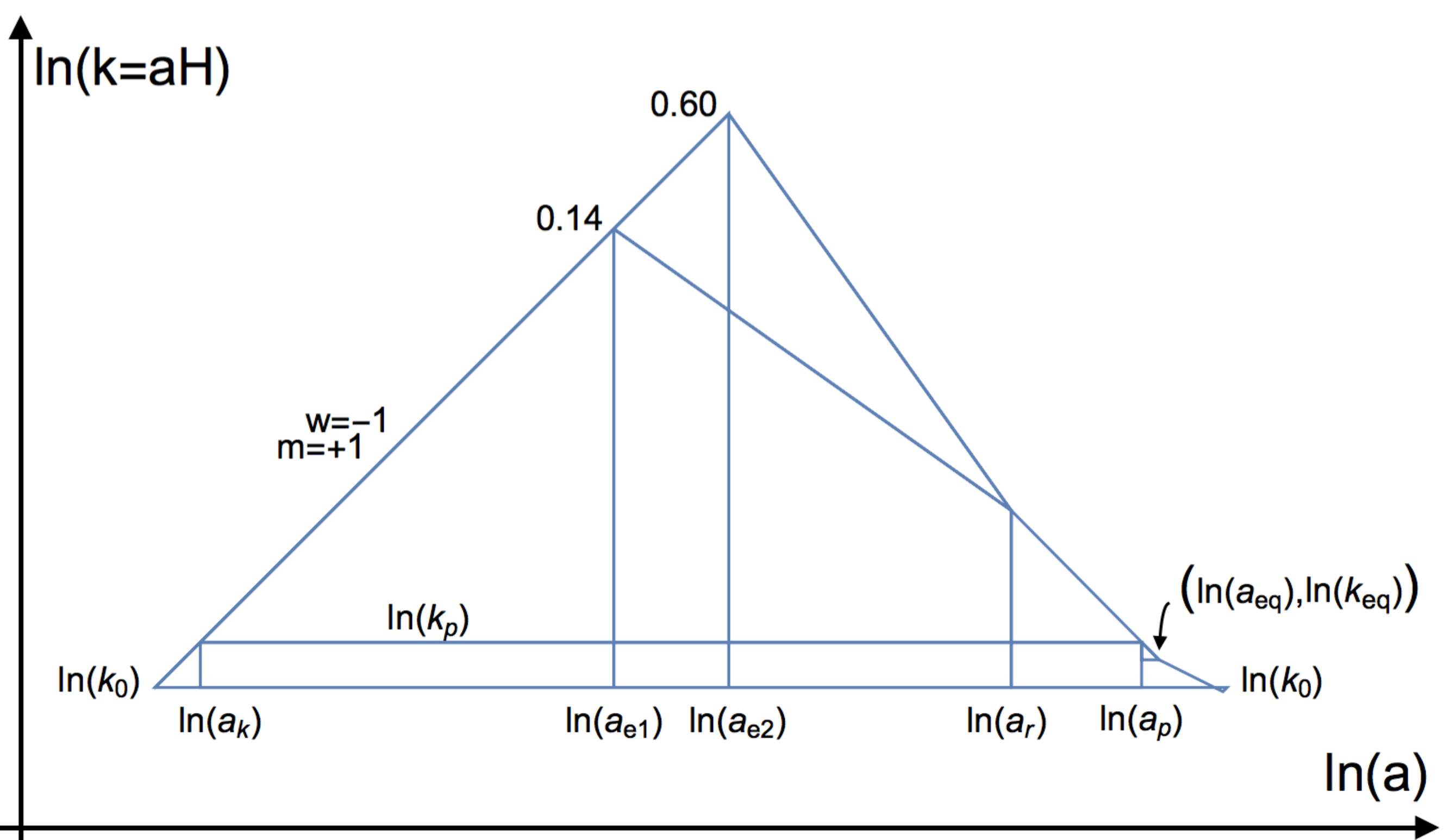}
\caption{\small Plot of the evolution lines during inflation, reheating and radiation for the bounds for $\omega_{re}$ given by Eq.~\eqref{boundsw} (the lines ending in the rounded up numbers 0.14 and 0.60) for the lowest possible reheating temperature of approximately $10\,MeV$. In Eq.~\eqref{boundsNT} the bounds are given, having been obtained by direct measurements from the figure as indicated by the projection of the lines on the $\ln(a)$ axis.
}
\label{DSbounds}
\end{figure}

To conclude this section we find the bounds when the minimum reheat temperature, around $10\,MeV$, is reached. Thus, we solve Eq.~\eqref{w} for $\omega_{re}$ when $T_{re}=10\,MeV$ using the bounds $0.9607 < n_s < 0.9691$ given by Planck. The EoS is bounded as (see Fig.~\ref{DSbounds})
\beq
\label{boundsw}
0.1390<\omega_{re}<0.6030,
\eeq
from where it follows that
\beq
\label{boundsNT}
48.86 < N_k < 62.53,   \quad 47.70 > N_{re} > 33.79,   \quad   N_{rd} =16.69,
\eeq
at $k=k_p$. We see, in particular, that the $\omega_{re}=0$ is excluded i.e., we cannot have very low reheating temperature for oscillations at the bottom of the Starobinsky potential. Note that the minimum value for $N_{rd}$ is fixed since it only depends on $T_{re}$.
%%%%%%%%%%%%%%%%%%%%%%%%%%%%%%%%%%%%%%%%%
%%%%%%%%%%%%%%%%%%%%%%%%%%%%%%%%%%%%%%%%%
%%%%%%%%%%%%%%%%%%%%%%%%%%%%%%%%%%%%%%%%%
\section {\bf Instantaneous reheating}\label{INS}
We now study the instantaneous reheating case because it is of some interest and because we can do an exact calculation. We have seen in the discussion related to Fig.~\ref{DSparalelas} that for $any$ EoS the limiting case of instantaneous reheating occurs when the number of e-folds during reheating vanishes with maximum reheating temperature. In this case the diagram representing the situation is perfectly symmetric from the scale of wavenumber mode $k=k_p$ upwards (upwards from horizontal $\ln(k_p)$ line in Fig.~\ref{DSparalelas}), with the radiation and the inflation lines making a right angle. 
This implies that, in this case, the number of e-folds from the end of inflation to the pivot scale is the same as the number of e-folds during inflation   thus, $N_{ep}=N_{ke}$. For the Starobinsky model the solution to the equation $N_{re}=0$ in the instantaneous reheating case is $\phi_k=5.364581...$, from where it follows that the tensor-to-scalar ratio is  $r=0.00343$, spectral index $n_s=0.9653$, running $\alpha=6.1\times 10^{-4}$, scale of inflation $\Delta=7.8\times 10^{15}GeV$, Hubble function $H_k=1.45\times 10^{13}GeV$, Hubble function at the end of inflation $H_e=9.65\times 10^{12}GeV$, reheating temperature $T_{re}=2.6\times 10^{15}\,GeV$, number of e-folds during inflation  $N_{ke}=55.6$ and number of e-folds during the radiation epoch $N_{rd}=57.7$ with $N_{ep}=55.5$ for the number of e-folds from $a_e$ to $a_p$. Note that $N_{ep}$ is not exactly the same as $N_{ke}$ as in the diagrammatic approach (again, the assumption of de Sitter inflation by the diagram) however, the difference is only 0.1 e-fold.

These results are the same for $any$ EoS including $\omega_{re}=1/3$. The $\omega_{re}=1/3$ case is interesting because analytically when $\omega_{re}=1/3$ there is no equation for $N_{re}$ \cite{Cook:2015vqa}. Diagrammatically the $\omega_{re}=1/3$ case is represented by the line of slope $m=-1$ joining the inflation and radiation lines. This does not mean, however, that there is no reheating and that radiation starts from the very tip at $V$ with instantaneous reheating; that is not what defines radiation. Reheating surely is a very complicated process to describe, starting perhaps with some non-perturbative particle production of an stage called
preheating (for a recent reference see \cite{Antusch:2020iyq} and references therein) 
and settling in the canonical scenario with the scalar field oscillating around the minimum of the potential. The scalar field, dominating the energy density of the universe up to the end of inflation, starts dissipating its energy by interactions with other fields and decaying into new particles while oscillations occur until a thermalization temperature is reached. After the inflaton has mostly or completely decayed it is when the radiation epoch starts: a very different era where relativistic particles, as the universe cools down, start decaying without any possibility of being created a new.
As our diagrammatic approach shows there is no reason why reheating cannot proceed with  $\omega_{re}=1/3$ as in all the other cases (all the other EoS). The only peculiarity of this case is that the line of reheating and the line of radiation have the same slope (same EoS). In the diagram it is just one line representing two very different processes.
In any case, for $\omega_{re}=1/3$ the number of e-folds during reheating is bounded as $0 < N_{re} < 40.91$ with $57.6 > N_{rd} > 16.69$ and $2.6\times 10^{15}\,GeV > T_{re} > 0.01\,GeV$ with all the other quantities as in the instantaneous reheating case discussed above.
%%%%%%%%%%%%%%%%%%%%%%%%%%%%%%%
%%%%%%%%%%%%%%%%%%%%%%%%%%%%%%%
\section {\bf Conclusions}\label{CON}
We have proposed an equation which allows to calculate the value of the inflation $\phi_k$ at the pivot scale of wavenumber mode $k_p$, this is given by Eq.~\eqref{EQ}. This equation is model independent in the sense that no model of inflation has been used to obtain it. However, its $solutions$ require the specification of a model through the Hubble function $H_k$ and the number of e-folds during inflation, denoted $N_{ke}\equiv \ln(\frac{a_e}{a_k})$, for a given $N_{ep}$. The determination of all inflationary observable and quantities of interest follows from $\phi_k$ (and from model dependent parameters, if any).  In the instantaneous reheating case we have that $N_{ep}=N_{ke}$ and Eq.~\eqref{EQ} can be solved exactly for any EoS.  We have established a clear connection between the analytical and diagrammatic methods and the simultaneous use of both of them can give a better comprehension of the phenomena described.
We have illustrated our diagrammatic approach in the $\omega_{re}=0$ case with the Starobinsky model of inflation as a concrete example. The diagrammatic approach can describe model independent situations while the analytical method always require the specification of a model of inflation. It is clear that the diagrammatic and analytical approaches complement each other and can be studied together  for better results. The strategy presented here can be applied to any model where $H_k $ and $N_{ke} $ can be obtained.
%%%%%%%%%%%%%%%%%%%%%%%%%%%%%%%
%%%%%%%%%%%%%%%%%%%%%%%%%%%%%%%
\section*{Acknowledgements}
We would like to thank Jaume Haro, Juan Carlos Hidalgo and Ariadna Montiel for discussions. It is also a pleasure to thank N. S\'anchez and M. Dirzo for encouraging conversations. We acknowledge financial support from UNAM-PAPIIT,  IN104119, {\it Estudios en gravitaci\'on y cosmolog\'ia}.
 %%%%%%%%%%%%%%%%%%%%%%%%%%%%%%%%
%%%%%%%%%%%%%%%%%%%%%%%%%%%%%%%%

\end{document}